\documentclass[a4paper]{article}

\usepackage[french,english]{babel}
\usepackage[T1]{fontenc}
\usepackage[utf8]{inputenc}
\usepackage{amsmath,amssymb,amsfonts,amsthm}
\usepackage{makeidx}
\usepackage{color}
\usepackage{empheq}
\usepackage{cancel}
\usepackage[top=20mm, bottom=20mm, left=20mm, right=20mm]{geometry}
\usepackage{stmaryrd} 
\usepackage{enumerate}
\usepackage{hyperref} 
\usepackage{booktabs}
\usepackage[all]{xy} 

\usepackage[vcentermath,noautoscale]{youngtab}
\everymath{\Yboxdim10pt}
\everydisplay{\Yboxdim10pt}

\SetSymbolFont{stmry}{bold}{U}{stmry}{m}{n} 

\theoremstyle{remark}

\theoremstyle{definition}

\newcommand{\Nb}{\mathbb{N}}

\newcommand{\Rr}{\mathrm{R}}
\newcommand{\Rb}{\mathbb{R}}

\newcommand{\Sc}{\mathcal{S}}
\newcommand{\Tc}{\mathcal{T}}
\newcommand{\Vc}{\mathcal{V}}

\renewcommand{\d}{\ensuremath{\operatorname{d}\!}}

\title{\textsf{\textbf{On the number of terms in the Lovelock products}}}
\author{\textsf{Xavier Lachaume}}
\date{\textsf{\normalsize{Summer 2017}}}

\makeindex

\begin{document}
\maketitle

\begin{center}
\textsf{
Institut Denis Poisson \\
Université de Tours - Université d'Orléans - UMR 7013 du CNRS \\
Parc de Grandmont - 37200 Tours - France \\
\texttt{xavier.lachaume@lmpt.univ-tours.fr}
}
\end{center}

\renewcommand*{\proofname}{$\textstyle\square \scriptstyle\square \scriptscriptstyle\square \diamond$}
\renewcommand{\qedsymbol}{$. \scriptscriptstyle \diamond \square \scriptstyle\square \textstyle\square$}

\vfill

\textsc{Abstract:}
In this short note we wonder about the explicit expression of the expanding of the $p$-th Lovelock product. We use the 1990s' works of S. A. Fulling et al. on the symmetries of the Riemann tensor, and we show that the number of independent scalars appearing in this expanding is equal to the number of Young diagrams with all row lengths even in the decomposition of the $p$-th plethysm of the Young diagram representing the symmetries of the Riemann tensor.

\vfill



\section{Introduction}

Lovelock theories are a set of modified gravity theories that can be seen as generalisations of General Relativity (GR) in higher dimension. They could have interesting cosmological implications (see \cite{Nojiri05GBDE}, \cite{Nojiri05MGBDE}, \cite{Nojiri06}, \cite{Cognola06}) or connections with string/M-theories in which higher-order curvature terms appear naturally (see \cite{Tor08}).

They can be represented in the form of an action by a sum of scalar contractions of multiple copies of the Riemann curvature tensor. The contraction of $p$ copies is called the $p$-th Lovelock product. In even dimension, the non-vanishing term of highest-degree coincides to the Gauss-Bonnet-Chern scalar of the space-time manifold, hence exhibits a promising relation with geometry.

Usually the Lovelock products are written as a product, and are handled in this form. This product is expanded only for small degrees, like $p = 1$, $2$ or $3$. The astonishing complexity of the expanding for $p=3$ in comparison to $p = 1$ or $2$ discourages to continue the expanding for further degrees. However the question is worth asking: what is the explicit formulation for the expanding of the general $p$-th Lovelock product? The only terms appearing in such a development are the Riemann tensor, the Ricci tensor and the scalar curvature of the space-time. But how many are they? And how are they combined? This is the topic of the present note. We do not solve entirely the problem, but give an answer to one of the questions: the number of independent scalars appearing in the expanding of the $p$-th Lovelock product.

\vspace{\baselineskip}

The answer we bring was in fact contained in a twenty-five years old paper from S. A. Fulling, R. C. King, B. G. Wybourne and C. J. Cummins, \cite{Ful92}. This field of research finds its origin in computational aspects of the heat kernel in the context of quantum field theory and gravitation. Algorithms to simplify tensor calculus were developed, such as for the Computer Algebra System (CAS) \texttt{REDUCE} (see \cite{Belkov96}) or \texttt{Mathematica} (see \cite{Booth98}).

Apparently we are the first one to make an explicit connection with Lovelock theories.

\section{Notations}

We represent the space-time by a lorentzian manifold $(\Vc,g_{\mu\nu})$ of dimension $n+1$, $n \in \Nb$ standing for the spatial dimension. We choose $c=\kappa=1$ for unit and $(-1,+n)$ for the signature of $g_{\mu\nu}$. We note
\begin{center}
\begin{tabular}{rp{8cm}}
	\toprule
$D$ & the Levi-Civita connection of $(\Vc,g_{\mu\nu})$, \\
$\Rr_{\mu\nu\rho\sigma}$ & the Riemann tensor of $D$, \\
$\Rr_{\mu\rho}$ & the Ricci tensor of $D$, \\
$\Rr$ & the curvature scalar of $D$, \\
$\d v = \sqrt{-g}\d^{n+1} x$ & the volume element of $\Vc$. \\
	\bottomrule
\end{tabular}
\end{center}
In its twice covariant and twice contravariant form, using its symmetries, the Riemann tensor can be written \[\Rr_{\gamma \delta}^{\alpha \beta} = \Rr_{\gamma \delta}^{\phantom{\gamma \delta} \alpha \beta} = \Rr_{\phantom{\alpha \beta} \gamma \delta}^{\alpha \beta}\] as well.

We introduce $p_n = \left\lfloor \dfrac{n+1}{2} \right\rfloor$ and
\[\delta_{\mu_1 \mu_2\ldots \mu_k}^{\nu_1 \nu_2\ldots \nu_k} := \det
		\begin{pmatrix}
		\delta_{\mu_1}^{\nu_1} & \ldots & \delta_{\mu_1}^{\nu_k} \\
		\vdots						 &				&	\vdots \\
		\delta_{\mu_k}^{\nu_1} & \ldots & \delta_{\mu_k}^{\nu_k}
		\end{pmatrix}\]
the generalised Kronecker symbol. We define
\begin{center}
\begin{tabular}{ll}
	\toprule
$\Rb_p = \dfrac{1}{2^p}\delta_{\alpha_1 \beta_1 \alpha_2 \beta_2 \cdots \alpha_p \beta_p}^{\gamma_1 \delta_1 \gamma_2 \delta_2 \cdots \gamma_p \delta_p}\Rr_{\gamma_1 \delta_1}^{\alpha_1 \beta_1}\Rr_{\gamma_2 \delta_2}^{\alpha_2 \beta_2} \ldots \Rr_{\gamma_p \delta_p}^{\alpha_p \beta_p}$ & the $p$-th Lovelock product. \\
	\midrule
$\Rb_0 = 1$, \\
$\Rb_1 = \Rr$ & is the scalar curvature, \\
$\Rb_2 = \Rr^2 - 4 \Rr_\alpha^\gamma\Rr^\alpha_\gamma + \Rr_{\alpha\beta}^{\gamma\delta}\Rr^{\alpha\beta}_{\gamma\delta}$ & corresponds to the Gauss-Bonnet term for $n+1=4$, \\
$\Rb_3 =
	\Rr^3
	+2\Rr_{\alpha\beta}^{\gamma\delta}\Rr_{\gamma\delta}^{\varepsilon\eta}\Rr_{\varepsilon\eta}^{\alpha\beta}
	+3\Rr\Rr_{\alpha\beta}^{\gamma\delta}\Rr_{\gamma\delta}^{\alpha\beta}$ \\
	\qquad $+8\Rr_{\alpha\beta}^{\gamma\eta}\Rr_{\gamma\delta}^{\varepsilon\beta}\Rr_{\varepsilon\eta}^{\alpha\delta} -12\Rr\Rr_\alpha^\beta\Rr_\beta^\alpha +16\Rr_\alpha^\beta\Rr_\beta^\gamma\Rr_\gamma^\alpha$ \\
	\qquad $-24\Rr_\alpha^\beta\Rr_{\gamma\delta}^{\alpha\varepsilon}\Rr_{\beta\varepsilon}^{\gamma\delta}
	+24\Rr_\alpha^\beta\Rr_\gamma^\delta\Rr_{\beta\delta}^{\alpha\gamma}$ & and so on, until \\
$\Rb_p = 0$ for $p > p_n$, & because of the antisymmetries of $\Rr_{\alpha\beta}^{\gamma\delta}$. \\
	\bottomrule
\end{tabular}
\end{center}

Then the action of a Lovelock theory is
\begin{equation}
\Sc_{\text{Lov}}[g] = \int_\Vc \sum_{p=0}^{p_n} \lambda_p \Rb_p \d v,
\end{equation}
with $\lambda_p$ real constants. In the following we shall focus on each $\Rb_p$.

\section{Young diagrams}

Even though a computer can deal with this computation, the quick explosion of the number of terms puts the question of an explicit formula for the expanding of the $\Rb_p$'s.

First of all, what are the terms involved in such a decomposition? If we write $\Sc_n$ the symmetric group on a set of $n$ elements, we get the formula:
\begin{align*}
\Rb_{p}
	&= \dfrac{1}{2^p}\delta_{\alpha_1 \beta_1 \ldots \alpha_p \beta_p}^{\gamma_1 \delta_1 \ldots \gamma_p \delta_p} \Rr_{\gamma_1 \delta_1}^{\alpha_1 \beta_1} \ldots \Rr_{\gamma_p \delta_p}^{\alpha_p \beta_p} \\
	&= \dfrac{1}{2^p} \sum_{\sigma \in \Sc_{2p}} \varepsilon(\sigma) \Rr_{\mu_1 \phantom{{}_{\sigma()}} \mu_2}^{\mu_{\sigma(1)} \mu_{\sigma(2)}} \ldots \Rr_{\mu_{2p-1} \phantom{{}_{\sigma()}} \mu_{2p}}^{\mu_{\sigma(2p-1)} \mu_{\sigma(2p)}}.
\end{align*}
This sum is on $(2p)!$ terms, which are obviously not linearly independent. The extraction of a basis among these terms is a problem which has been solved in \cite{Ful92}, using tools of the group representation theory which we shall shortly present here.

For all $k \geq 1$, every $k$-tensor $T_{a_1 \ldots a_k}$ can be mapped to a representation of $\Sc_k$. Just as this representation can be decomposed into irreducible representations of $\Sc_k$ encoded by Young diagrams of size $k$, the tensor $T_{a_1 \ldots a_k}$ can be decomposed as well on a basis $\Tc_k$ of $k$-tensors corresponding to Young diagrams of size $k$. More precisely, the tensors of $\Tc_k$ have peculiar symmetries which are encoded in standard Young tableaux of size $k$. We note with $\longleftrightarrow$ this correspondence.

Let us explain how to proceed to such a decomposition: for each tableau of size $k$, take $T_{a_1 \ldots a_k}$, symmetrise on the indices lying in each row, and then antisymmetrise on the indices lying in each column. For instance,
\begin{align*}
\young(13,2)\cdot T_{abc} &:= \dfrac{1}{3}\left(\left[T_{abc} + T_{cba}\right] - \left[T_{bac} + T_{cab}\right]\right), \\
\young(1,2,3)\cdot T_{abc} &:= \dfrac{1}{6}\left(T_{abc} + T_{bca} + T_{cab} - T_{bac} - T_{acb} - T_{cba}\right), \\
\end{align*}
where the combinatorial factors $\frac{1}{3}=\frac{2}{6}$ and $\frac{1}{6}$ are the normalised numbers of standard Young tableaux of the same Young diagram. The normalisation constant is the sum on the diagrams of the squared number of standard tableaux of this diagram, which turns out to be $k!$. For example, for $k=3$,
\[\begin{array}{ll}
\text{Diagrams} & \text{Tableaux} \\
\toprule
\yng(3) & \young(123) \\
\midrule
\yng(2,1) & \young(12,3) \quad \young(13,2) \\
\midrule
\yng(1,1,1) & \young(1,2,3) \\
\bottomrule
\end{array}\]
and we find $1^2+2^2+1^2 = 3!$. Thus each 3-tensor can be decomposed into:
\[T_{abc} = \left(\yng(3) + \yng(2,1) + \yng(1,1,1)\right)\cdot T_{abc} = \left(\young(123) + \young(12,3) + \young(13,2) + \young(1,2,3)\right)\cdot T_{abc}.\]

The product of two tensors $U_{a_1 \ldots a_k} \in \Tc_k$ and $U'_{b_1 \ldots b_l} \in \Tc_l$, whose symmetries correspond to the diagrams $\mathfrak{U}$ and $\mathfrak{U}'$, corresponds itself to the outer product of $\mathfrak{U}$ and $\mathfrak{U'}$, namely $\mathfrak{U}\cdot\mathfrak{U}'$. This product can in turn be decomposed into irreducible representations of $\Sc_{k+l}$, following the Littlewood-Richardson rule. For example, if $U_{abc}$ and $U'_{abc}$ are 3-tensors represented by
\[U_{abc} \longleftrightarrow \young(13,2) \text{\qquad and \qquad} U'_{abc} \longleftrightarrow \young(12,3),\]
then $\mathfrak{U} = \mathfrak{U}' = \yng(2,1)$, and
\begin{align*}
U_{abc}U'_{def}
	&= \left(\yng(2,1)\cdot\yng(2,1)\right)\cdot U_{abc}U'_{def} \\
	&= \left(\Yboxdim6pt\yng(4,2) + \yng(4,1,1) + \yng(3,3) + 2\ \yng(3,2,1) + \yng(3,1,1,1) + \yng(2,2,2) + \yng(2,2,1,1) \right)\cdot U_{abc}U'_{def}.
\end{align*}
An interesting theorem proved in \cite{Ful92} states that among these irreducible representations of $\Sc_{k+l}$, those of which all rows have an even length are invariant under the orthogonal group $O(k+l)$. Moreover, they form a basis of these invariant representations. In terms of tensors on a manifold, the invariant representations under the action of $O(k+l)$ correspond to the scalars built from the contractions of $U_{a_1 \ldots a_k}$ and $U'_{b_1 \ldots b_l}$. Hence, the space of the scalars built from the contractions of $U_{a_1 \ldots a_k}$ and $U'_{b_1 \ldots b_l}$ has the same dimension as the space of the irreducible representations of $S_{k+l}$ which are invariant under $O(k+l)$ and which appear in the decomposition of $\mathfrak{U}\cdot\mathfrak{U}'$. So, a basis of independent scalars will have the same cardinal as the number of diagrams with even row length in the decomposition of $\mathfrak{U}\cdot\mathfrak{U}'$.

Intuitively, contracting a pair of indices of the tensors can be understood as crossing a pair of cells off the Young diagram. If the two cells lie in different rows, the result vanishes because of the antisymmetrisation between the rows. If the two cells lie in the same row, the contraction is nontrivial. Hence, all rows must have even lengths so that the resulting empty diagram correspond to a nontrivial scalar.

However, the equality of the dimensions does not imply a simple bijective correspondence between the even row length diagrams and the independent scalars: most of the time, such a correspondence does not exist. Keeping our example, we have two even row length representations:
\[\yng(4,2) \text{\qquad and \qquad} \yng(2,2,2),\]
and two independent scalars built from the contractions of $U_{a_1 \ldots a_k}$ and $U'_{b_1 \ldots b_l}$:
\[U_{ab}^{\phantom{ab}a} {U'}_c^{\phantom{c}cb} \text{\qquad and \qquad} U_{abc}{U'}^{cab},\]
but no canonical bijection between them.

In case $U_{a_1 \ldots a_k} = U'_{b_1 \ldots b_l}$, the list of irreducible representations is restricted by the symmetries under the exchange of the two tensors: this is not an outer product anymore, but a new operation called a \textbf{plethysm}, $\otimes$. This is the case we are interested in.

\section{Plethysms of the Riemann tensor}

Indeed, our aim is to determine a basis of scalars on which the $\Rb_p$'s can be decomposed. Now, each $\Rb_p$ is a sum of all the possible contractions of $p$ Riemann tensors. If we notice that the Riemann tensor has symmetries verifying:
\[\Rr_{\mu_1\mu_2}^{\mu_3\mu_4} \longleftrightarrow \young(13,24),\]
we can conclude that we have to study the decomposition of
\[\yng(2,2)\otimes\yng(2,2)\otimes\ldots\otimes\yng(2,2) =: \yng(2,2)^{\otimes p}\]
onto irreducible representations. In this decomposition, the number of even row length diagrams will be the number of independent scalars which can be made by contractions of $p$ copies of $\Rr_{\mu\nu}^{\rho\sigma}$. For instance, keeping only the even row length diagrams, we get
\[\Yboxdim8pt \begin{array}{llccc}
\toprule
\Rb_1 \longleftrightarrow & \yng(2,2)^{\otimes 1} = & \yng(2,2) \\
& & \downarrow \\
& & \Rr \\
\midrule
\Rb_2 \longleftrightarrow & \yng(2,2)^{\otimes 2} = & \Yboxdim6pt\yng(4,4) &
\begin{array}{rl} +&\Yboxdim6pt\yng(4,2,2) \end{array} &
\begin{array}{rl} +&\Yboxdim6pt\yng(2,2,2,2) \end{array} \\
& & \downarrow & \downarrow & \downarrow \\
& & \Rr^2 & \Rr_{\alpha\beta}\Rr^{\alpha\beta} & \Rr_{\alpha\beta}^{\gamma\delta}\Rr^{\alpha\beta}_{\gamma\delta} \\
\midrule
\Rb_3 \longleftrightarrow & \yng(2,2)^{\otimes 3} = & \Yboxdim4pt\yng(6,6) &
\begin{array}{rl} +&\Yboxdim4pt\yng(6,4,2) \\ \\ +&\Yboxdim4pt\yng(4,4,4) \end{array} &
\underbrace{\begin{array}{rlrl} +&\Yboxdim4pt\yng(6,2,2,2) & +&\Yboxdim4pt\yng(4,4,2,2) \\ \\ +&\Yboxdim4pt\yng(4,4,2,2) & +&\Yboxdim4pt\yng(4,2,2,2,2) \\ \\ +&\Yboxdim4pt\yng(2,2,2,2,2,2) \end{array}} \\
& & \downarrow & \downarrow & \downarrow \\
& & \\
& & \Rr^3 & \Rr\Rr_\alpha^\beta\Rr_\beta^\alpha & \Rr_{\alpha\beta}^{\gamma\delta}\Rr_{\gamma\delta}^{\varepsilon\eta}\Rr_{\varepsilon\eta}^{\alpha\beta} \quad \Rr\Rr_{\alpha\beta}^{\gamma\delta}\Rr_{\gamma\delta}^{\alpha\beta} \\
& & & \Rr_\alpha^\beta\Rr_\beta^\gamma\Rr_\gamma^\alpha & \Rr_{\alpha\beta}^{\gamma\eta}\Rr_{\gamma\delta}^{\varepsilon\beta}\Rr_{\varepsilon\eta}^{\alpha\delta} \quad \Rr_\alpha^\beta\Rr_{\gamma\delta}^{\alpha\varepsilon}\Rr_{\beta\varepsilon}^{\gamma\delta} \\
& & & & \Rr_\alpha^\beta\Rr_\gamma^\delta\Rr_{\beta\delta}^{\alpha\gamma} \\
\bottomrule
\end{array}\]
As we explained, there is no one-to-one correspondence between the diagrams and the scalars. However, it is possible to sort them in three subsets which are bijectively connected, as we did in the previous array. If the diagram contains:
\begin{itemize}
\item[$\star$] two rows, it represents a scalar involving only the scalar curvature $\Rr$;
\item[$\star$] three rows, it represents a scalar involving $\Rr$ and the Ricci tensor $\Rr_{\mu\nu}$;
\item[$\star$] four rows or more, it represents a scalar involving $\Rr$, $\Rr_{\mu\nu}$ and the Riemann tensor $\Rr_{\mu\nu}^{\rho\sigma}$.
\end{itemize}

Some tables were computed in \cite{Ful92}, which allow us to count the number of scalars for the first $\Rb_p$'s:
\[\begin{array}{ccccccc}
\text{Order} && \multicolumn{3}{c}{\text{Number of rows}} && \text{Total} \\
\toprule
&& 2 & 3 & \geq 4 && \\
\midrule
\Rb_1 && 1 & 0 & 0 && 1 \\
\Rb_2 && 1 & 1 & 1 && 3 \\
\Rb_3 && 1 & 2 & 5 && 8 \\
\Rb_4 && 1 & 3 & 22 && 26 \\
\Rb_5 && 1 & 4 & 85 && 90 \\
\Rb_6 && 1 & 6 & 402 && 409 \\
\bottomrule
\end{array}\]
Unfortunately, these numbers were calculated by a computer; there is no explicit formula to determine them. As well for the explicit form of the scalars.

A fortiori in the computation of $\Rb_p$. After having summed upon all the permutations of $\Sc_{2p}$:
\[\sum_{\sigma \in \Sc_{2p}} \varepsilon(\sigma) \Rr_{\mu_1 \phantom{{}_{\sigma()}} \mu_2}^{\mu_{\sigma(1)} \mu_{\sigma(2)}} \ldots \Rr_{\mu_{2p-1} \phantom{{}_{\sigma()}} \mu_{2p}}^{\mu_{\sigma(2p-1)} \mu_{\sigma(2p)}},\]
there is currently no explicit formula about the factors in front of each of these scalars. All we know is that their sum is $(2p)!$, and that the symmetries ensure that all factors can be divided by $2^p$.

For the rest, there is no formula. Yet there exist algorithms able to deal with the computation. One can find such algorithms and applications for \texttt{Maple} in \cite{Portugal98}, \cite{Portugal99}, \cite{Portugal01}, \cite{Portugal02}. There are also algorithms for the language \texttt{REDUCE}, \cite{Ilyin96}, or \texttt{Java}, \cite{Ilyin00}. Algorithms for tensor simplification in \texttt{Mathematica} (package Tools of Tensor Calculus) can be found in \cite{Balfagon98}, \cite{Balfagon99}. The program \texttt{Cadabra} can be used as well.

\section{Conclusion}

We give an answer to the question of the number of independent scalars in the expanding of the $p$-th Lovelock product: it is equal to the number of Young diagrams with all row lengths even in the decomposition of the $p$-th plethysm of the Young diagram $(2,2)$ that encodes the symmetries of the Riemann tensor:
\[\yng(2,2)^{\otimes p}.\]

The following questions are still open: can one find an explicit formula for this number? Afterwards, how many each of those scalars are present in the expanding
\[\delta_{\alpha_1 \beta_1 \ldots \alpha_p \beta_p}^{\gamma_1 \delta_1 \ldots \gamma_p \delta_p} \Rr_{\gamma_1 \delta_1}^{\alpha_1 \beta_1} \ldots \Rr_{\gamma_p \delta_p}^{\alpha_p \beta_p}?\]
We only know that their sum is $(2p)!$, and the symmetries ensure that the number of their appearances is a multiple of $2^p$. In small dimensions, this is not a big deal. In higher-dimensional theories however, eg. a string theory with $n+1=26$, the number of independent scalars as well as their factors explode.

\section*{Acknowledgments}
I would like to thanks Professor Stephen Fulling for having made himself available and for the useful references he gave to me.

\bibliographystyle{unsrt}
\bibliography{../../Biblio}

\end{document}